\documentclass{PoS}
\usepackage{array,multirow}
\usepackage{pgfplots}
\usepackage{graphicx}
\usepackage{epsfig}
\usepackage{amsfonts}
\usepackage{amsmath}
\usepackage{amssymb}
\usepackage{dsfont}
\usepackage{color}
\usepackage{cite}
\usepackage{pdflscape}
\usepackage{pifont}

    \renewcommand{\arraystretch}{1.5}

\newcommand{\al}{\alpha}
\newcommand{\bt}{\beta}
\newcommand{\g}{\gamma}

\newcommand{\simu}{\sigma^{\mu\nu}}

\newcommand{\Or}{\mathcal O}

\newcommand{\vL}{\ensuremath{\mathcal{L}}}
\newcommand{\vp}{\varphi}
\newcommand{\sq}{^{2}}

\newcommand{\dslash}[1]{#1 \llap{/\kern-0.5pt}}
\newcommand{\Dslash}[1]{#1 \llap{/\kern+1.5pt}}
\newcommand{\DDslash}[1]{#1 \llap{/\kern+2.3pt}}
\newcommand{\dslashh}[1]{#1 \llap{/\kern+1pt}}

\newcommand{\bea}{\begin{eqnarray}}
\newcommand{\eea}{\end{eqnarray}}
\newcommand{\bma}{\begin{pmatrix}}
\newcommand{\ema}{\end{pmatrix}}
\newcommand{\nn}{\nonumber}

\title{High- and low-energy constraints on top-Higgs couplings}

\ShortTitle{Constraining top-Higgs couplings}

\author{\speaker{W.\ Dekens}\\
        Los Alamos National Laboratory, New Mexico Consortium\\
        E-mail: \email{wdekens@newmexicoconsortium.org}}

\abstract{
We study the five chirality-flipping interactions that appear in the top-Higgs sector at leading order in the standard model effective field theory. We consider constraints from collider observables, flavor physics, and electric-dipole-moment experiments. This analysis results in very competitive constraints from  indirect observables when one considers a single coupling at a time. 
In addition, we discuss  how these limits are affected in scenarios in which multiple top-Higgs interactions are generated at the scale of new physics.}

\FullConference{9th International Workshop on the CKM Unitarity Triangle\\
		28  November - 3 December 2016\\
		Tata Institute for Fundamental Research (TIFR), Mumbai, India}

\begin{document}
\section{Introduction}
As the top quark has the largest coupling to the Higgs boson, and thereby the electroweak symmetry breaking sector, it might be the most sensitive to influences of new physics. This is exemplified in several beyond-the-standard-model (BSM) scenarios relevant for baryogenesis \cite{Kaplan:1991dc,Agashe:2006wa,Carena:2008vj,Kobakhidze:2014gqa}. In these cases one expects enhanced deviations from the SM to occur in the interactions of the top quark and the Higgs boson. These top-Higgs couplings can be probed directly by measurements of processes involving top quarks at the LHC. Other, `indirect', constraints come from processes which do not involve the top quark, but to which the top-Higgs couplings can contribute through loop corrections. As we will discuss, such indirect observables can give rise to complementary constraints, and, in some cases, are more stringent than the direct limits.

In the following we discuss direct and indirect limits on dimension-six chirality-flipping top-Higgs interactions within the framework of the standard model effective field theory (SM-EFT).  
We do not consider chirality conserving top-Higgs operators,
 for recent analyses see e.g.\ \cite{Hartmann:2016pil,Buckley:2015lku,RHCCpaper}, 
which only mix with the top-Higgs couplings at three loops \cite{Jenkins:2013zja,Jenkins:2013wua,Alonso:2013hga}. 
We introduce the chirality-flipping operators in section \ref{sec:operators}. An overview of their contributions to direct observabes (such as single-top, $t\bar t$, and $t\bar t h$ production) and indirect collider observables ($gg\leftrightarrow h$, $h\to \g\g$) is given in section \ref{sec:collider}. Observables in flavor physics (such as $B\to X_s\g$) and electric dipole moment (EDM) searches are discussed in sections \ref{sec:flavor} and \ref{sec:EDM}. We consider the resulting limits, and discuss how these constraints are affected when all couplings are turned on at the same time, in section  \ref{sec:discussion}.

\section{Operator structure}\label{sec:operators}
Assuming that new physics arises at a scale $\Lambda$ well above the electroweak scale, $\Lambda\gg v$ where $v\simeq 246$ GeV,  BSM effects can be described by higher-dimensional operators in the SM-EFT. Here we keep only effects linear in $v\sq/\Lambda\sq$ which can be described by the minimal set of dimension-six operators that has been derived in \cite{Buchmuller:1985jz,Grzadkowski:2008mf}. If we further assume that the dominant BSM effects appear in the chirality-flipping top-Higgs sector, the dimension-six Lagrangian involves five operators,
\begin{equation}
{\cal L}_{\rm top}   =
\sum_{\alpha \in \{ Y, g, \gamma, Wt,Wb\} }  \ C_\alpha  \, O_\alpha + {\rm h.c.} \qquad  \quad
\label{eq:Leff}
 \end{equation}
where  $C_\alpha = c_\alpha + i \,   \tilde{c}_\alpha$  are complex couplings. In the unitary gauge and in the quark mass basis the operators are given by,
\bea \label{eq:operators1}
O_\gamma &=&  - \frac{e Q_t}{2}  m_t  \, \bar{t}_L  \sigma_{\mu \nu} \left( F^{\mu \nu} - t_W  Z^{\mu \nu} \right) t_R \, \left(1 + \frac{h}{v}\right)   \,,
\qquad 
\\
O_g &=&  - \frac{g_s}{2}  m_t  \, \bar{t}_L  \sigma_{\mu \nu} G^{\mu \nu} t_R \, \left(1 + \frac{h}{v}\right)   \,,
\\
O_{Wt} \!\!   &=& -g m_t \bigg[  \frac{1}{\sqrt{2}}  \bar{b}_L'   \simu  t_R W_{\mu\nu}^- 
+  \bar t_L\simu t_R \bigg(\frac{1}{2c_W} Z_{\mu\nu}+i g W_\mu^-W_\nu^+\bigg)\bigg]\bigg(1+\frac{h}{v}\bigg) \,,\qquad
\\
O_{Wb} \!\! &=& - g m_b   \bigg[\frac{1}{\sqrt{2}} \bar t_L'  \simu b_R  W_{\mu\nu}^+ 
- \bar b_L\simu b_R \bigg(\frac{1}{2c_W} Z_{\mu\nu}+i g W_\mu^-W_\nu^+\bigg)\bigg]\bigg(1+\frac{h}{v}\bigg) \,,
\\
O_Y  &=&   -  m_t  \bar{t}_L  t_R  \left( v h  + \frac{3}{2}  h^2 + \frac{1}{2}  \frac{h^3}{v}  \right) \label{eq:operators2} ~, 
\eea
where $F_{\mu\nu},\,Z_{\mu\nu},\, W^\pm_{\mu\nu}$, and $G^a_{\mu\nu}$ are the field strengths of the photon, $Z$ boson,  $W^{\pm}$ boson, and gluon, while   $e,\, g$, and $g_s$ denote the $U(1)_{\rm EM}$, $SU(2)$ and $SU(3)_c$ gauge couplings, respectively. Furthermore, $h$ represents the Higgs field, $Q_t = 2/3$,  $t_W = \tan \theta_W$,  $c_W = \cos \theta_W$,  with $\theta_W$ the Weinberg angle. 
Finally, the operators $O_{Wt,Wb}$  contain the combinations  $b' = V_{tb} b + V_{ts} s + V_{td} d$, and  $t' = V_{tb}^* t + V_{cb}^* c + V_{ub}^* u$, where $V_{ij}$ represent the SM CKM elements.  
These operators, in the mass basis, can be related to the $SU(2)\times U(1)_Y$ invariant operators of \cite{Buchmuller:1985jz,Grzadkowski:2008mf} as discussed in Refs.\ \cite{Cirigliano:2016njn,Cirigliano:2016nyn}.
Furthermore, at low energies the first of the above operators can be interpreted as the electric and  magnetic dipole moments of the top quark, ($d_t = (e m_t Q_t)  \tilde{c}_\gamma$ and  $\mu_t = (e m_t Q_t)  {c}_\gamma$). The second is related to the non-abelian gluonic electric and  magnetic dipole moments, 
($\tilde d_t =  m_t  \tilde{c}_g$ and  $\tilde \mu_t =  m_t {c}_g$). 

In what follows we will assume that the  operators in Eqs.\ \eqref{eq:operators1} - \eqref{eq:operators2} capture the dominant BSM effects, and therefore set all other dimension-six operators to zero at the scale $\Lambda$. However, the renormalization group evolution (RGE), as well as threshold effects, will induce additional dimension-six operators. As we will see below, these additional operators will lead to stringent constraints from indirect observables.
\section{Collider observables and electroweak precision tests}\label{sec:collider}
The operators in Eqs.\ \eqref{eq:operators1} - \eqref{eq:operators2} can be probed directly in the production and decay of top quarks. Of the observables in the former category we consider single top \cite{Cirigliano:2016nyn}, $t\bar t $ \cite{Atwood:1994vm,Haberl:1995ek},  and $t\bar t h$ \cite{Degrande:2012gr,Hayreter:2013kba} production, while we study the $W$ helicity fractions in $t\to Wb$ decays \cite{Drobnak:2010ej}. 
These processes receive corrections from the top-Higgs couplings at tree level. In particular, single top production and the helicity fractions are sensitive to corrections to the $Wtb$ vertex generated by $C_{Wt}$. Both $t\bar t$ and $t\bar t h$ production receive corrections from the gluonic dipole moment, $C_g$, while the top Yukawa, $C_Y$ only affects $t\bar t h$. We summarize which of the observables receive contributions from the top-Higgs couplings in the left panel of Table \ref{Tab:Collider}.

At the one-loop level, the $C_\al$ couplings can also contribute to  processes without a top quark in the final state. Examples of such indirect  observables are electroweak precision tests, namely the $S$ parameter, and  Higgs production and decay signal strengths, in particular, $h\leftrightarrow gg$, and $h\to \g\g$. Both Higgs production and decay channels are induced at loop-level in the SM, such that the loop generated BSM contributions can be sizable in comparison. These contributions are induced by the following additional Higgs-gauge operators 
\bea\label{ColliderO}
O_{\vp\g}= e\sq F_{\mu\nu}F^{\mu\nu}\vp^\dagger \vp,\qquad O_{\vp G}=g_s\sq G^a_{\mu\nu}G^{a,\, \mu\nu}\vp^\dagger \vp,\qquad O_{\vp WB}=gg' W^I_{\mu\nu}B^{\mu\nu}\vp^\dagger \tau^I\vp\,
\eea
where $\vp$ is the Higgs doublet and $B_{\mu\nu}$ ($g'$) is the $U(1)_Y$ field strength (coupling constant), and $B_\mu = c_W A_\mu-s_W Z_\mu$. The $C_{\g}$ and $C_{g}$ couplings induce the $O_{\vp \g}$ and $O_{\vp \g}$ operators through RG evolution between $\mu=\Lambda$ and $\mu=m_H$ \cite{Jenkins:2013zja,Jenkins:2013wua,Alonso:2013hga}, while $C_Y$ induces both operators  through threshold effects. The additional operator $O_{\vp WB}$ is generated by $C_\g$ and $C_{Wt}$. In turn, $O_{\vp \g}$ and $O_{\vp G}$ contribute to $h\to\g\g$ and $h\leftrightarrow gg$, while $O_{\vp WB}$ induces the $S$ parameter. These contributions are summarized in the right panel of Table \ref{Tab:Collider}.

As we only consider effects linear in $v\sq/\Lambda\sq$, we do not take into account contributions that are quadratic in the top-Higgs couplings. This implies that measurements of cross sections are only sensitive to the real parts of $C_\al$. Of the above mentioned observables only  the phase $\delta^-$, which is measured in $t\to Wb$ decays \cite{Boudreau:2013yna,Aad:2015yem}, is sensitive to the imaginary part of $C_{Wt}$.

\begin{table}\centering\small
\renewcommand{\arraystretch}{1.5}
{\arraycolsep=4pt
$\begin{array}{c|c||c cccc}
&{\rm \textbf{Obs.}}& C_\g & C_g&C_{Wt}&C_{Wb}&C_Y
\\\hline\hline
&t& $\ding{55} $&$\ding{55} $ & \large \checkmark& $\ding{55} $&$\ding{55} $ \\ 
&
t\bar t& $\ding{55} $&\large\checkmark & $\ding{55} $ &$\ding{55} $ & $\ding{55} $\\\rotatebox{90}{\rlap{\textbf{Direct}}}
&t\bar t h& $\ding{55} $& \large\checkmark& $\ding{55} $& $\ding{55} $&\large \checkmark\\
 &t\to Wb& $\ding{55} $& $\ding{55} $&\large\checkmark & $\ding{55} $&$\ding{55} $\\
\end{array}$ }
\qquad \qquad 
\renewcommand{\arraystretch}{1.7}
$\begin{array}{c|c||c cccc}
&{\rm \textbf{Obs.}}& C_\g & C_g&C_{Wt}&C_{Wb}&C_Y
\\\hline\hline
&gg\leftrightarrow h& $\ding{55} $&O_{\vp G}&$\ding{55} $ &$\ding{55} $ & O_{\vp G}\\
\rotatebox{90}{\rlap{\hspace{-.7cm} \textbf{Indirect}}}
&h\to \g\g&\begin{tabular}{c}$O_{\vp\g}$\end{tabular}& $\ding{55} $& $\ding{55} $&$\ding{55} $ &O_{\vp\g}  \\\cline{2-7}& S& O_{\vp WB}  & $\ding{55} $&O_{\vp WB} &  $\ding{55} $&$\ding{55} $
\end{array}$
\caption{The left and right panels give an overview of the contributions of the top-Higgs couplings to direct and indirect  observables, respectively. A $\checkmark$ indicates a direct (tree-level) contribution and \ding{55} a negligible contribution. The $O_i$ indicate indirect contributions induced through the additional operator $O_i$ defined in Eq.\ \eqref{ColliderO}. 
}\label{Tab:Collider}
\end{table}

\section{Flavor physics}\label{sec:flavor}
At scales $\mu\simeq m_b$ the top-Higgs couplings give rise to (flavor-changing) interactions that can, in principle,  be probed in multiple flavor observables. Here we  focus on the flavor-changing $B\to X_s\g$ transitions which give rise to the most stringent constraints. We consider the branching ratio and the CP-asymmetry, both of which are induced by the dipole operators $O_7$ and $O_8$ \cite{Lunghi:2006hc,Kagan:1998ym,Benzke:2010tq},
\bea
O_7 = \frac{e}{16\pi\sq}m_b\bar s_L \simu F_{\mu\nu}b_R,\qquad O_8 = -\frac{g_s}{16\pi\sq}m_b\bar s_L \simu G^a_{\mu\nu}t^ab_R,
\eea
which appear in the Lagrangian, $\vL_{b\to s} = -4V_{tb}V_{ts}^*G_F/\sqrt{2}\sum_{i=7,8}C_i O_i$. These two operators are induced through the RG evolution from the scale $\Lambda$ to the $b$ quark mass scale. All top-Higgs operators, apart from the top Yukawa $C_Y$, contribute to these operators through one-loop diagrams, although the resulting constraints are most relevant for $C_{Wt}$ and $C_{Wb}$. The combination of the branching ratio and the CP asymmetry then leads to constraints on the real and imaginary parts of these top-Higgs couplings.

\section{Electric dipole moments}\label{sec:EDM}

EDMs are generated by additional operators that the top-Higgs couplings induce through loop effects. The set of additional operators that  give rise to EDMs at low energies, $\mu\sim 2$ GeV, is given by
\bea\label{eq:EDMs}
\vL_{\rm EDM} &=& -\frac{i}{2}\sum_{q=u,d,s,c,b}\bigg[e Q_q m_q\, C_\g^{(q)} \bar q \simu \g_5  q\,F_{\mu\nu}+g_s m_q\, C_g^{(q)} \bar q \simu \g_5 t^a  q\,G^a_{\mu\nu}\bigg]\nn\\
&&+C_{\tilde G}\frac{g_s}{3}f_{abc}\tilde G^{a\, \mu\nu}G^b_{\mu\rho}G^{c,\, \rho}_{\nu} -\frac{i}{2}e Q_e m_e\, C_\g^{(e)} \bar e \simu \g_5 e\,F_{\mu\nu} ,
\eea
where $\tilde G^a_{\mu\nu} = \frac{1}{2}\epsilon_{\mu\nu\al\bt}G^{a\, \al\bt}$. 
The operators in the first line are the quark EDMs and quark color-EDMs  and the first term in the second line represents the CP-odd three-gluon operator, all of which induce the EDMs of nucleons and nuclei. These EDMs can be calculated by first performing the nonperturbative matching to an extension of chiral perturbation theory that incorporates CPV hadronic interactions \cite{deVries:2012ab,Bsaisou:2014oka}. The EDMs of the  nucleons, $d_{n}$ and $d_p$, can then be expressed in terms of these CPV hadronic interactions, while for nuclear systems, such as $^{199}$Hg, nuclear-structure calculations are required to do so. 
Both the matching to  chiral effective theory  and the nuclear calculations involve significant uncertainties. The  theoretical errors  are under control in the case of the contributions of the quark EDMs to $d_{n,p}$, due to lattice calculations with $\Or(15\%)$ uncertainties \cite{Bhattacharya:2015esa,Bhattacharya:2015wna,Bhattacharya:2016zcn}, while calculations for the color-EDMs are underway \cite{Bhattacharya:2016rrc,Abramczyk:2017oxr}.
Moving on from the quark EDMs, other uncertainties range from  $\Or(50\%)$, for the contributions of the color-EDMs to $d_{n,p}$, to over $100\%$ with an unknown sign, in case of the dependence of $d_{\rm Hg}$ on the CPV pion-nucleon couplings. 
In contrast, the measurement of the ThO molecule, $d_{\rm ThO}$, has a rather clean theoretical interpretation in terms of the electron EDM (last operator in Eq.\ \eqref{eq:EDMs}). For reviews of these issues see e.g.\ \cite{Engel:2013lsa,Yamanaka:2017mef}.
Here we will  employ the EDM measurements of the neutron, ThO, and mercury \cite{Baker:2006ts,Baron:2013eja,Graner:2016ses}. We briefly comment on the impact of theory errors in section \ref{sec:discussion}, see  \cite{Chien:2015xha,Cirigliano:2016nyn} for details.

\begin{table}\centering\small
\renewcommand{\arraystretch}{1.5}
\arraycolsep=8pt
$\begin{array}{c||c cccc}
{\rm \textbf{Obs.}}& \tilde c_\g & \tilde c_g&\tilde c_{Wt}&\tilde c_{Wb}&\tilde c_Y
\\\hline\hline
d_{\rm ThO}& O_{\vp\tilde X,lequ}\to O_\g^{(e)} &O_\g\to O_{\vp\tilde X,lequ}\to O_\g^{(e)} & \large O_{\vp\tilde X,lequ}\to O_\g^{(e)}&$\ding{55}$  &O_{\g}^{(e)} \\
 
d_n,\, d_{\rm Hg}& O_{\vp\tilde X,quqd}\to O_{\g,\, g}^{(q)} &O_{\tilde G} & O_{\vp\tilde X,quqd}\to O_{\g,\, g}^{(q)} &O_g^{(b)}\to O_{\tilde G}  & O_{\g,\,g}^{(q)},\, O_{\tilde G}
\end{array}$ 
\caption{\small The Table gives an overview of the contributions of the top-Higgs couplings to EDMs. \ding{55} indicates a negligible contribution. The other entries 
illustrate the mechanism through which the dominant  contributions arise and the additional operators (defined in Eqs.\ \eqref{eq:EDMs} and \eqref{eq:extraEDM}) that are induced in the process.
}\label{Tab:EDMs}
\end{table}

To generate $d_{n}$, $d_{\rm ThO}$ and $d_{\rm Hg}$  one first has to induce the operators in  Eq.\ \eqref{eq:EDMs}. 
$C_Y$ generates all operators through two-loop Barr-Zee diagrams \cite{Barr:1990vd,Gunion:1990iv}, while $C_g$  ($C_{Wb}$) does so at  the one (two) loop level by inducing the   $O_{\tilde G}$ operator.  The operators $O_{\g,\, Wt}$ can also induce EDMs at one loop, however, such contributions are generated through a $W$ loop which involves a small factor of $|V_{ub}|\sq$. In addition, these diagrams only generate the quark (color-)EDMs and do not induce the, experimentally more stringently constrained, electron EDM. As a result,  a two-loop mechanism gives rise to limits on $O_{\g,\, Wt}$ which are stronger by roughly three orders of magnitude  \cite{Cirigliano:2016njn,Cirigliano:2016nyn}. In this mechanism,  one-loop RG evolution first induces additional operators, which are not present in Eq.\ \eqref{eq:EDMs}, namely,
\bea\label{eq:extraEDM}
\vL_{\rm EDM}' 
&=&\vp^\dagger\bigg[C_{\vp\tilde B}g^{\prime\, 2}  B^{\mu\nu}\tilde B_{\mu\nu} +C_{\vp\tilde W}g^{ 2}  W^{I\,\mu\nu}\tilde W^I_{\mu\nu}+C_{\vp\tilde WB}gg^{\prime} B^{\mu\nu}\tau\cdot \tilde W_{\mu\nu} +C_{\vp\tilde G}g_s^{ 2}  G^{a\,\mu\nu}\tilde G^a_{\mu\nu}\bigg]\vp \nn\\
&&+C_{lequ} (\bar l^I_L\simu e_R) \epsilon_{IJ} (\bar q_L^J \sigma_{\mu\nu}u_R)\nn\\
&&+C_{quqd}^{(1)} (\bar q^I_L\simu u_R )\epsilon_{IJ} (\bar q_L^J \sigma_{\mu\nu}d_R)+C^{(8)}_{quqd} (\bar q^I_L\simu t^a u_R) \epsilon_{IJ} (\bar q_L^J \sigma_{\mu\nu}t^ad_R).
\eea
In the second step a subset of the above operators, $C_{\vp \tilde WB,\vp \tilde W,\vp \tilde B,lequ}$, generates the electron EDM through an additional loop. In similar fashion, $C_{\vp \tilde G}$ and $C_{quqd}^{(1,8)}$ generate the quark (color-)EDMs and thereby hadronic EDMs. 
Since all of the operators in Eq.\ \eqref{eq:extraEDM} are induced by $O_{\g,\, Wt}$, the stronger experimental  limit on the electron EDM gives the most stringent bounds on these top-Higgs couplings.
The contributions of all top-Higgs couplings, and the mechanisms by which they do so, are summarized in Table \ref{Tab:EDMs}.

\section{Discussion}\label{sec:discussion}
\begin{figure}[t!]
\center
\includegraphics[width=9cm]{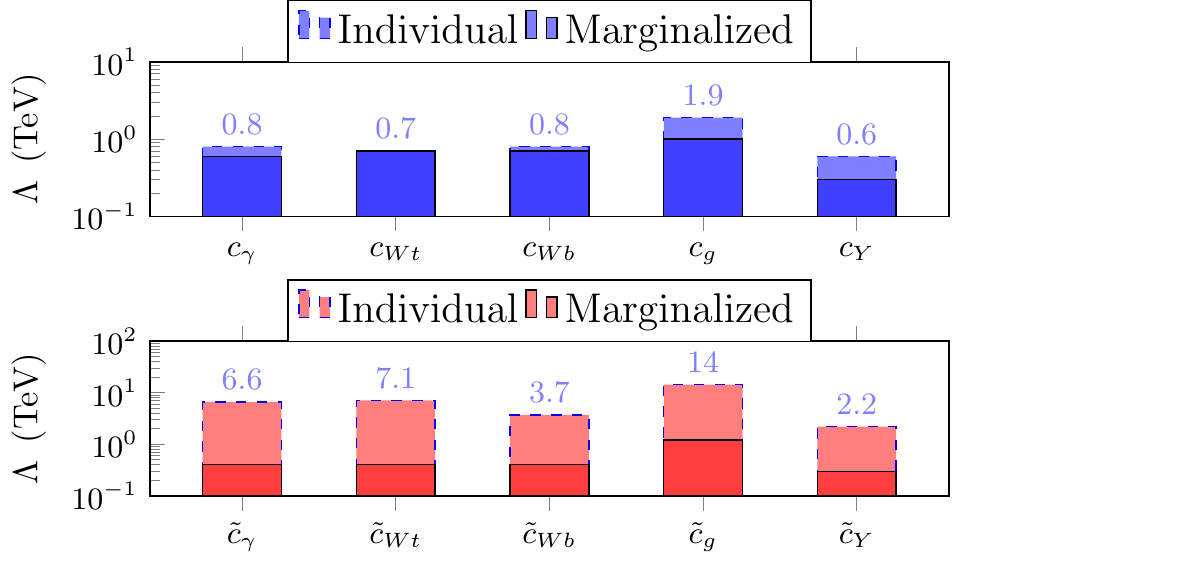}\hspace{-.5cm}
\includegraphics[width=4.5cm]{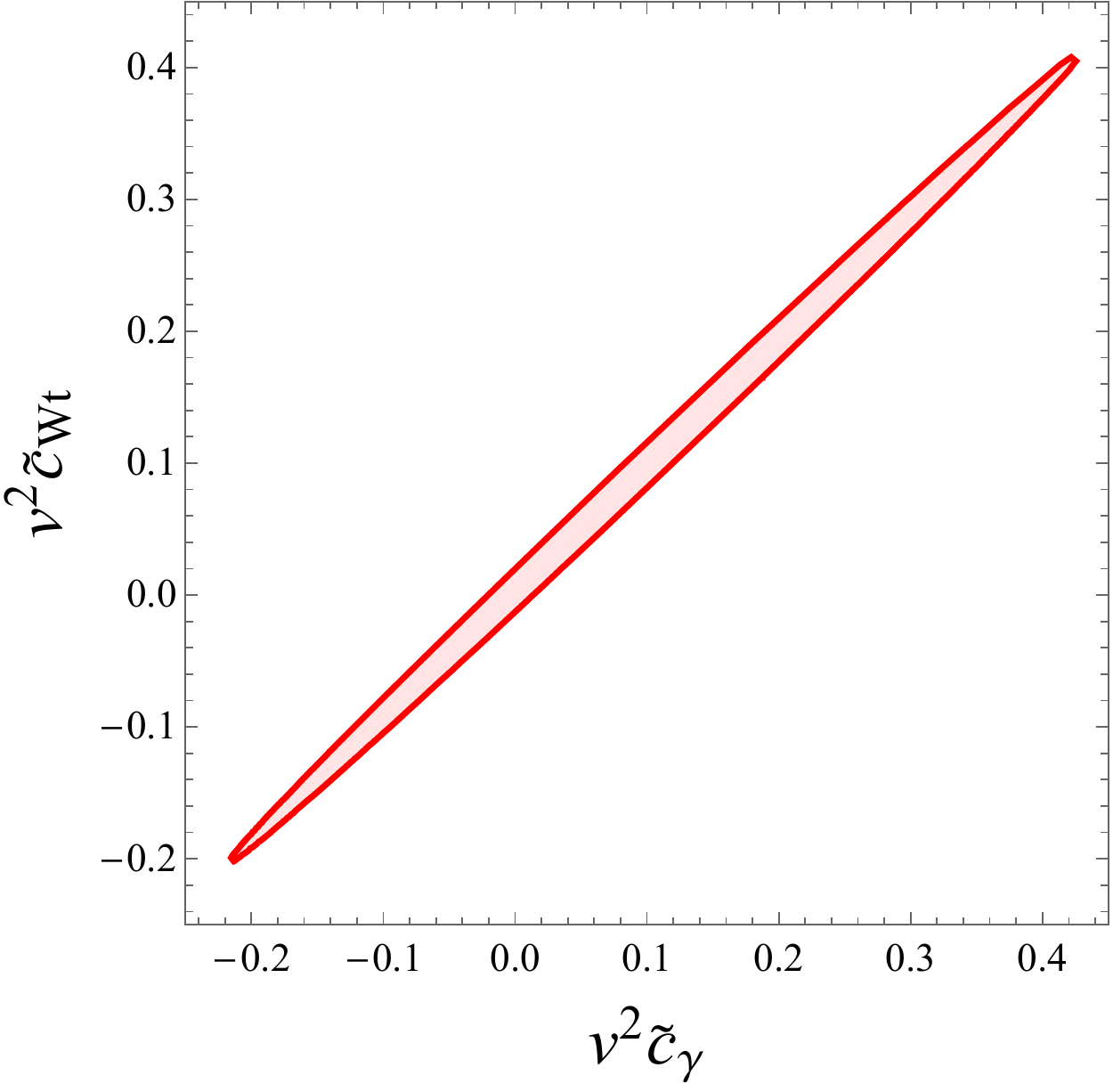}
\caption{\small The solid  bars in the left panel show the constraints on each operator, $C_\al$, after marginalizing over the other operators. The dashed bars give the limits assuming only one operator is present. The limits are naively translated to a scale of new physics by using $C_\al= v\sq/\Lambda\sq$. The right panel shows the limits in the $v\sq \,\tilde c_\g- v\sq\, \tilde c_{Wt}$ plane after marginalizing over the other  couplings. }
\label{charts}
\end{figure}

We summarize the constraints that result from the above outlined analysis in Fig.\ \ref{charts}. Here the top and bottom panels on the left show the limits on the real and imaginary parts, respectively. The dashed bars indicate the bounds in the scenario that a single top-Higgs coupling is turned on at the high scale. One can see that the limits on the imaginary parts are generally stronger than those on the real parts (note the different scales in Fig.\ \ref{charts}). In this scenario the limits on the real part of  $C_{Wb}$ comes  from $B\to X_s\g$, while the limits on $C_\g$, $C_g$, and $C_Y$ arise mainly  from the indirect collider observables $h\to\g\g$ and $gg\leftrightarrow h$. The bound on the real part of $C_{Wt}$ is dominated by the direct observables in $t\to Wb$. 
The limits on all imaginary parts are dominated by EDMs, $C_{\g,Wt,Y}$ are stringently constrained by $d_{\rm ThO}$ while the stronger constraints on $C_{g,Wb}$ come from $d_{n}$ and $d_{\rm Hg}$. The constraints resulting from $d_{\rm ThO}$ are hardly affected by the theory uncertainties mentioned in section \ref{sec:EDM}, while those from $d_n$ and $d_{\rm Hg}$ weaken significantly. The limit on $C_g$ ($C_{Wb}$) is weakened by  two (one) order of magnitude if one uses the `Rfit' approach \cite{Charles:2004jd} for these uncertainties.

The solid bars in Fig. \ref{charts} give the limits in the scenario when all five operators are present at the scale $\Lambda$. As can be seen from the left panel, the real parts are mildly affected, while the limits on the imaginary parts are drastically weakened. This is due to significant cancellations between different contributions to the EDMs. As an example of this we show the $\tilde c_\g- \tilde c_{Wt}$ plane in the right panel of Fig.\ \ref{charts}. Here the ThO measurement limits the couplings to the diagonal band. However, this indirect observable alone allows for a free direction, which is only ruled out by direct probes (the helicity fractions). This further motivates  collider observables which are directly sensitive to CP violation such as the ones proposed in \cite{Prasath:2014mfa,Boudjema:2015nda,Mileo:2016mxg}.

\section*{Acknowledgments}
I would like to thank the organizers of the CKM2016 workshop,  in particular the conveners of Working Group 6, for an interesting and enjoyable meeting. I am grateful to Vincenzo Cirigliano, Emmanuele Mereghetti, and Jordy de Vries for the collaboration on this work. This work was supported by the Dutch Organization for Scientific Research (NWO) through a RUBICON grant.

\bibliographystyle{h-physrev3} 
\bibliography{bibliography}

\end{document}